# Description of $^{160}Dy$ nucleus by Partial Dynamical $SU(3)$ Symmetry

N. Fouladi[a], M. A. Jafarizadeh[b,c1], J. Fouladi[a], H. Sabri[a2]

[a] Department of Nuclear Physics, University of Tabriz, Tabriz 51664, Iran.
[b] Department of Theoretical Physics and Astrophysics, University of Tabriz, Tabriz 51664, Iran.
[c] Research Institute for Fundamental Sciences, Tabriz 51664, Iran.

---

[1] E-mail: jafarizadeh@tabrizu.ac.ir
[2] E-mail: h-sabri@tabrizu.ac.ir

**Abstract**

We considered the characteristic features of $SU(3)$ partial dynamical symmetry in the interacting boson model framework to demonstrate the relevance of this technique in the nuclear spectroscopy of $^{160}Dy$ nucleus. The predictions of $SU(3)-PDS$ for energy spectrum and $B(E2)$ transition probabilities were compared with the most recent experimental data which an acceptable degree of agreement is achieved.

During the last three decades, dynamical symmetries (DS) have been used extensively in many complex systems and have led to many important discoveries in diverse areas of physics with notable examples in nuclear, molecular, hadronic, polymer and nanostructure physics [1-5]. Dynamical symmetries can be viewed as a generalization and refinement of the exact symmetry concepts. Recently, the Partial Dynamical Symmetry (PDS) has been introduced [7-11] in order to a further enlargement of the fundamental concepts of the exact and dynamical symmetries. Partial dynamical symmetry provides an intermediate symmetry structure which corresponds to a particular symmetry breaking, but preserves the useful aspects of a dynamical symmetry for a part of the system. The advantage of using interactions with a PDS is that they can be introduced, in a controlled manner, without destroying results previously obtained with a DS for a segment of the spectrum [11]. One important aspect of PDS is their ability to serve as a practical tool for calculation of observables in realistic systems.

The mathematical aspects and algorithm for constructing the Hamiltonians with partial dynamical symmetry has been developed in Ref.[7] and further elaborated to Hamiltonians with higher-order terms presented in Ref.[10] by Leviatan *et al*. Also, the relevance of $SU(3)-PDS$ to the spectroscopy of $^{168}Er$ nucleus have been described in Ref.[8] and showed that, this nucleus can be a good example of $SU(3)-PDS$, i.e. the resulting PDS calculations are found to be in excellent agreement with experimental data. The purpose of this paper is to show that PDS are capable to investigate and analyze the spectral properties of nuclear systems. We consider the $^{160}Dy$ nucleus as a typical example of axially deformed prolate nuclei in the rare earth region [12] and show the relevance of $SU(3)-PDS$ to its description.

The $SU(3)-DS$ is an appropriate symmetry structure introduced in the interacting boson model (IBM) framework for describing the axially deformed nuclei which based on the pioneering works of Elliott [6]. IBM [13-15] provides a rich algebraic structure to illustrate the implications of the partial dynamical symmetry which is widely used in description of low-lying collective states in nuclei. Therefore, we consider the relevant aspects of this model which related to the $SU(3)-PDS$. The IBM description of an axially deformed nucleus is the $SU(3)$ limit which describes a symmetric rotor with degenerate $\beta$ and $\gamma$ bands. The basis states in this limit are labeled by $|[N](\lambda,\mu)KLM\rangle$ where $N$ is the total number of bosons, $(\lambda,\mu)$ denote the $SU(3)$ irreducible representations (irreps), $L$ is the angular momentum and $K$ is the multiplicity label.

This extra quantum number, i.e. $K$, which corresponds geometrically to the projection of the angular momentum on the symmetry axis, is necessary for complete classification. Each $K$-value in a given $SU(3)$ irrep $(\lambda,\mu)$, is associated with a rotational band and in different $K$-bands, states with the same $L$ are degenerate. The ground $g(K=0)$ band of an axially deformed prolate nucleus which described by irrep $(2N,0)$, is the lowest $SU(3)$ irrep. On the other hand, both the $\beta(K=0_2^+)$ and $\gamma(K=2_1^+)$ bands, which is used to describe the lowest excited bands, span the irrep $(2N-4,2)$ and therefore, the states in $\beta$ and $\gamma$ bands with the same $L$ are degenerate. This undesired $\beta-\gamma$ degeneracy, which is a characteristic feature of the $SU(3)$ limit in IBM framework, can be lifted by adding an extra term from other chains to the $SU(3)$ Hamiltonian, although this kind of $K$-band degeneracy is not commonly observed in deformed nuclei [12]. In the empirical spectra of most deformed nuclei the $\beta$ and $\gamma$ bands are not degenerate and thus, to conform the experimental data, one is compelled to break $SU(3)$ symmetry. Such an $SU(3)$ symmetry breaking introduced by Warner, Casten and Davidson (WCD) [14] or similar approach was taken in the consistent Q formalism (CQF) by the same authors [15] in order to lift the undesired $\beta-\gamma$ degeneracy. In these procedures, where an additional term from other chains was added to the $SU(3)$ Hamiltonian, the $SU(3)$ symmetry is completely broken, all eigenstates are mixed and none of virtues are retained. In contrast, Leviatan [8], have introduced the partial dynamical $SU(3)$ symmetry in which corresponds to a particular $SU(3)$ symmetry breaking, but preserves the useful aspects of a dynamical symmetry, e.g., the solvability for a part of the system. Hamiltonian of $SU(3)-DS$ composed of a linear combination of the Casimir operators of $SU(3)$ and $O(3)$ groups. A two-body $SU(3)$-PDS Hamiltonian in the framework of IBM has the form [11]

$$\hat{H}_{PDS} = \hat{H}(h_0,h_2) + C\hat{C}_{O(3)} = h_0 P_0^\dagger P_0 + h_2 P_2^\dagger \cdot \tilde{P}_2 + C\hat{C}_{O(3)} \qquad , \qquad (1)$$

Where $\hat{H}(h_0,h_2)$ is a two-body Hamiltonian with partial $SU(3)$ symmetry, $P_0^\dagger = d^\dagger \cdot d^\dagger - 2(s^\dagger)^2$ and $P_{2\mu}^\dagger = 2d_\mu^\dagger s^\dagger + \sqrt{7}(d^\dagger d^\dagger)_\mu^{(2)}$ are the boson-pair operators in IBM with angular momentum $L=0$ and $2$, respectively and $\hat{C}_{O(3)}$ denotes the Casimir operator of $O(3)$ group. For $h_0 = h_2$ case, the $\hat{H}(h_0,h_2)$ form an $SU(3)$ scalar related to the Casimir operator of $SU(3)$ while for $h_0 = -5h_2$, it is an $SU(3)$ tensor, namely $(\lambda,\mu)=(2,2)$. Although the $\hat{H}(h_0,h_2)$ is not a $SU(3)$ scalar, it has a subset of solvable states with good $SU(3)$ symmetry. The additional $O(3)$ rotational term which converts the partial $SU(3)$ symmetry into $SU(3)-PDS$, contributes an $L(L+1)$ splitting and has no effect on wave functions and consequently, the undesired $\beta-\gamma$ degeneracy can be lifted. According to the prescription introduced in Ref.[11], the solvable states of $\hat{H}_{PDS}$ which preserve the $SU(3)$ symmetry, are members of the ground $g(K=0)$ and $\gamma^k(K=2k)$ bands and have the form

For $g(K=0): |N,(2N,0),K=0,L\rangle$ and therefore $\Rightarrow E_{PDS}=CL(L+1), L=0,2,...,2N$ (2a)

For $\gamma^k(K=2k):$ $|N,(2N-4k,2k),K=2k,L\rangle$ and therefore $\Rightarrow$

$E_{PDS}=6h_2k(2N-2k+1)+CL(L+1),$ $L=K,K+1,...,(2N-2k)$ (2b)

$\hat{H}_{PDS}$, i.e. Eq.(1), is specified by three parameters, namely $C, h_2$ and $h_0$, which the values of $C$ and $h_2$ were extracted from the experimental energy spectra and the parameter $h_0$ was varied so as to reproduce the bandhead energy of the $\beta$ band.

To determine these quantities for our considered nucleus, $^{160}Dy$ with $N=14$, we have employed the same method introduced in Refs.[8,11]. With employing the latest empirical data taken from Refs.[17-18], we extracted, $h_0=8.7, h_2=5.4$ and $C=14.4\ kev$. The resultant spectra by the predictions of the $SU(3)-PDS$, the corresponding experimental spectrum and also the predictions of $SU(3)-DS$ for $^{160}Dy$ nucleus are presented in Figure1.

As have explained extensively in Refs.[8-11], the experimental spectrum of deformed nuclei and especially, the $\beta(K=0_2^+)$ and $\gamma(K=2_1^+)$ bands are not degenerate. On the other hand, one can expect, the spectrum of an exact $SU(3)-DS$ which obtained by $h_0=h_2$ and therefore indicate degeneracy of these bands deviates considerably from the empirical data. The lifted $\beta-\gamma$ degeneracy governed by the predictions of $SU(3)-PDS$, show an improvement over the schematic description of exact $SU(3)$ dynamical symmetry. Similar suggestions, namely a closer corresponding between the predictions of $SU(3)-PDS$ with empirical spectra for considered nucleus is apparent in the Figure 1.

Also, as it is evident from the experimental spectrum of $^{160}Dy$, in the most of deformed nuclei, the $\beta$ band lies above the $\gamma$ band which confirmed by the predictions of $SU(3)-PDS$ spectrum.

The reduced electric quadrupole transition probabilities as well as quadrupole moment ratios within the low-lying state bands, are regard as most striking investigation tools to consider the predictions of different models about the structure of the states. Therefore, we can employ these observables as a significant indicator to conform the predictions of $SU(3)-PDS$ about the considered nucleus. The $E2$ transition operator must be a Hermitian tensor of rank two and consequently the number of bosons must be conserved. The most general one body $E2$ operator in the IBM framework, is given by [11]

$T(E2)=\alpha Q^{(2)}+\theta\ \Pi^{(2)}$ , (3)

Where $Q^{(2)}=(d^\dagger s+s^\dagger\tilde{d})-\frac{\sqrt{7}}{2}(d^\dagger\tilde{d})_\mu^{(2)}$ is the quadrupole $SU(3)$ generator and $\Pi^{(2)}=(d^\dagger s+s^\dagger\tilde{d})$ is a $(2,2)$ tensor under $SU(3)$. The matrix elements of such $E2$ operator are determined by Lachello *et al* [13] and Van Isacker [16]. Also, the analytic expressions of $B(E2)$ values for intraband transitions, i.e. $g\rightarrow g$ and interband transitions, namely $\beta\rightarrow g$ and $\gamma\rightarrow g$, have been derived in Refs.[13,16] by the same authors.

The relative $B(E2)$ values thus depend on the parameters $\alpha$ and $\theta$ which can be extracted from the experimental values taken from Refs.[17-18]. To determine these quantities, we have employed the prescription introduced in Refs.[8,11] which terminate the corresponding ratio, i.e. $\theta/\alpha = 3.098$ for $^{160}Dy$ nucleus.

In Table 1, we compare the predictions of $SU(3)-PDS$ and IBM (which have determined by the same method introduced in Refs.[13-14]) for the $B(E2)$ branching ratios from different states in the $g, \beta$ and $\gamma$ bands of $^{160}Dy$ nucleus with the corresponding experimental values.

Table1. A comparison between the predictions of PDS and IBM for $B(E2)$ ratios [W.u.] with the experimental counterparts taken from Refs.[17-18].

| $J_i^\pi \quad J_f^\pi$ | EXP | PDS | DS | $J_i^\pi \quad J_f^\pi$ | EXP | PDS | DS |
|---|---|---|---|---|---|---|---|
| $2_g^+ \to 0_g^+$ | 305.46 | 288.33 | 269.21 | $2_\gamma^+ \to 0_g^+$ | 2.20 | 2.18 | 2.12 |
| $4_g^+ \to 2_g^+$ | 94.13 | 97.23 | 104.05 | $2_\gamma^+ \to 2_g^+$ | 4.23 | 4.29 | 4.41 |
| $6_g^+ \to 4_g^+$ | 67.36 | 64.50 | 75.08 | $2_\gamma^+ \to 4_g^+$ | 0.29 | 0.27 | 0.22 |
| $8_g^+ \to 6_g^+$ | 89.64 | 81.26 | 79.11 | $2_\beta^+ \to 0_g^+$ | 0.51 | 0.43 | 0.39 |
| $10_g^+ \to 8_g^+$ | 88.65 | 82.04 | 95.04 | $2_\beta^+ \to 2_g^+$ | 0.75 | 0.65 | 0.71 |
| $12_g^+ \to 10_g^+$ | 83.79 | 85.74 | 91.06 | $2_\beta^+ \to 4_g^+$ | 1.27 | 1.52 | 1.55 |

From this table, we see a closer corresponding between the predictions of $SU(3)-PDS$ for different $B(E2)$ values and their experimental counterparts. In particular, the calculated $B(E2)$ ratios in the $SU(3)-PDS$ framework for $\gamma \to g$ transitions lead to the parameter-free predictions which are in the excellent agreement with the corresponding experimental values.

From these Figures and Tables, one can conclude, the determined results indicate the elegance of the fits presented in this technique and they suggest the success of the estimation processes. Since, the Partial dynamical symmetry lifts the remaining degeneracy between $\beta$ and $\gamma$ bands but preserves the symmetry of the selected states, therefore, the acceptable degree of agreement between the predictions of this approach and the experimental counterparts, confirm the relevance of $SU(3)-PDS$ to the spectroscopy of $^{160}Dy$.

In summary, we considered the energy levels and $B(E2)$ transition probabilities of $^{160}Dy$ nucleus in the $SU(3)-PDS$ framework. The validity of the presented parameters, i.e. $h_0, h_2$ and $C$, has been investigated and it is seen that there is an existence of a satisfactory agreement between the presented results and experimental data. We may conclude that the general characteristics of the considered nucleus are well accounted in this study and the idea of the lifted $\beta-\gamma$ degeneracy by $SU(3)-PDS$ for this nucleus, is supported. The elegance of Figures1 and Table1 suggest an acceptable agreement between the presented $SU(3)-PDS$ results and the experimental data for considered nucleus. The obtained results in this study confirm that this technique is worth extending for investigating the nuclear structure of other nuclei existing in this mass region.

**Figure caption**

**Figure1(color online).** The energy spectra of $^{160}Dy$. Experimental energies (EXP) are compared with the IBM predictions of exact $SU(3)$ dynamical symmetry [SU(3)] and partial dynamical $SU(3)$ symmetry (PDS). The latter employed the $SU(3) - PDS$ Hamiltonian introduced in Eq.(1) with $h_0 = 8.7$, $h_2 = 5.4$ and $C = 14.4 kev$.

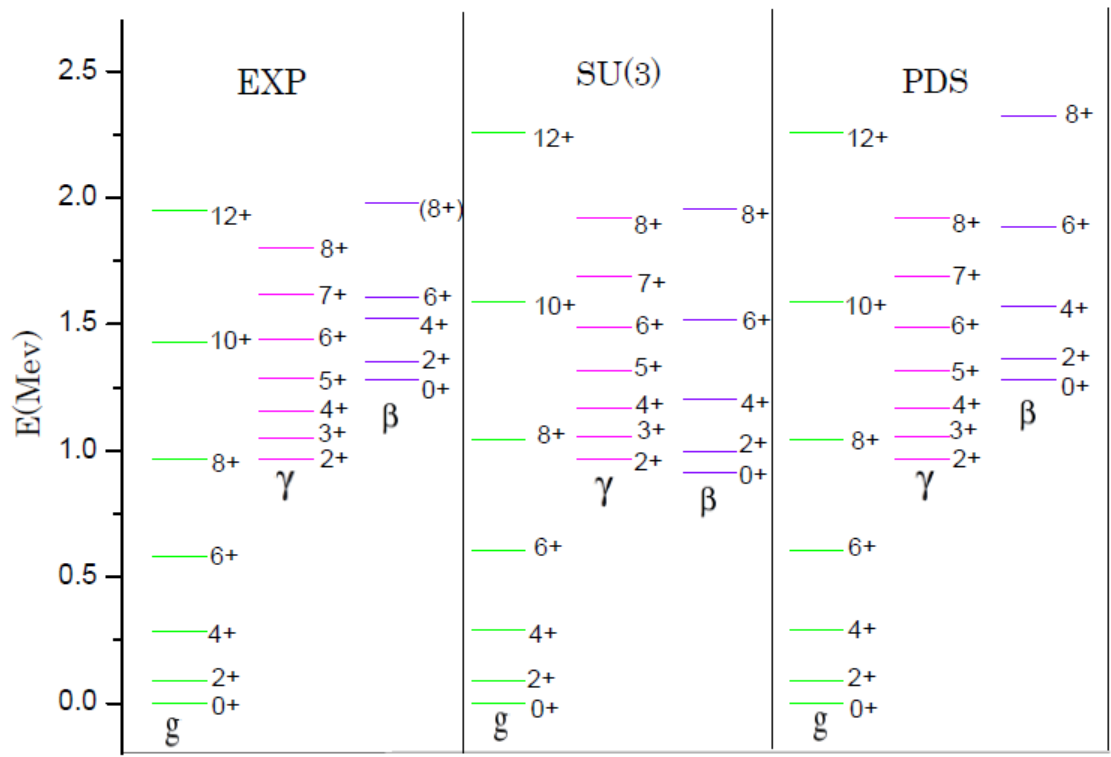